\documentclass[12pt,preprint]{aastex}
\usepackage{natbib}

\def\ergcm2s{erg cm$^{-2}$ s$^{-1}$} 
\def\etal{et al.}		
\def\n4038{NGC 4038/39}		

\def\ergcm2s{~erg cm$^{-2}$ s$^{-1}$ } 
\def\etal{et al.~}              

\def\n4038{~NGC4038/39}         

\def\x2{$\chi^{2}$}     

\begin{document}

\title{{\it Chandra}'s Discovery of Activity in the Quiescent Nuclear Black Hole
 of NGC~821
}

\author{G. Fabbiano$^1$, A. Baldi$^1$, S. Pellegrini$^2$, A. Siemiginowska$^1$, M. 
Elvis$^1$, A. Zezas$^1$, J. McDowell$^1$}
\affil{$^1$Harvard-Smithsonian Center for Astrophysics,
60 Garden Street, Cambridge, MA 02138; gfabbiano@cfa.harvard.edu; 
abaldi@cfa.harvard.edu, aneta@cfa.harvard.edu, elvis@cfa.harvard.edu,  
azezas@cfa.harvard.edu, jcm@cfa.harvard.edu}
\affil{$^2$Dipartimento di Astronomia, Universita` di Bologna, via Ranzani 1, 40127 
Bologna (Italy); silvia.pellegrini@unibo.it}

\shorttitle{XXXXXX}
\shortauthors{Fabbiano et al.}
\bigskip

\date{\sc May 18, 2004}
\bigskip

\begin{abstract}
We report the results of the {\it Chandra} ACIS-S observations of the
elliptical galaxy NGC~821, which harbors a supermassive nuclear black
hole (of $3.5 \times 10^7 M_{\odot}$), but does not show sign of AGN
activity.  A small, 8.5$^{\prime\prime}$ long ($\sim 1$~kpc at the
galaxy's distance of 23~Mpc), S-shaped, jet-like feature centered on
the nucleus is detected in the 38~ksec ACIS-S integrated exposure of
this region. The luminosity of this feature is $L_X \sim 2.6 \times
10^{39} \rm ergs~s^{-1}$ (0.3-10~keV), and its spectrum is hard
(described by a power-law of $\Gamma = 1.8^{+0.7}_{-0.6}$; or by
thermal emission with $kT >2$~keV).  We discuss two possibilities for
the origin of this feature: (1) a low-luminosity X-ray jet, or (2) a hot
shocked gas. In either case, it is a clear indication of nuclear
activity, detectable only in the X-ray band. Steady spherical accretion of
the mass losses from the central stellar cusp within the 
accretion radius, when coupled to a high radiative efficiency, already
provides a power source exceeding the observed radiative losses from
the nuclear region.

\end{abstract}

\section{Introduction}

Luminous quasars, radio galaxies, and Seyfert galaxies have long been
associated with accretion onto a massive black hole (see review Rees
1984), and the lack of nuclear emission in most galaxies was
alternatively debated as evidence for the lack of such a nuclear black
hole, or for the lack of fuel reaching the hole (Phinney 1983).

We now know that virtually all galaxies host supermassive black holes
(SMBH) in their nuclei.  High resolution observations of the nuclei of
elliptical galaxies and bulges have established the presence of these
SMBHs (Richstone et al 1998; Magorrian et al 1998; van der Marel
1999), whose mass is loosely correlated with the galaxy/bulge
luminosity (Magorrian et al 1998) and tightly correlated with the
central velocity dispersion (Ferrarese \& Merritt 2000; Gebhardt et al
2000) and the central light concentration (Graham et al 2001).  These
results remove the black hole variable from the equation, and make the
absence of luminous AGN in these galaxies more puzzling.  Although a
few faint AGNs have been detected in X-rays (e.g, in the $L_X/L_E \sim
10^{-(6-7)}$ range, where $L_E$ is the Eddington luminosity 
of the SMBH; Fabbiano \& Juda 1997; Pellegrini et al. 2003a; Ho
et al. 2001; Fabbiano et al. 2003), we still do not have a clear picture
of what impedes the formation of a luminous AGN, and of how this is
related to the existence and physics of a circum-nuclear accretion disk
(e.g., Martini et al. 2001).  Given that elliptical galaxies tend to
host large quantities of centrally concentrated hot interstellar
medium (ISM;  see Fabbiano 1989 for an early review), 
lack of fuel does not seem to be an option.

For example, the circum-nuclear regions studied so far with the
{\it Chandra} ACIS show the presence of hot gas close to the accretion
radius; this implies, when applying the spherical Bondi
accretion theory, accretion luminosities values comparable to those of
luminous AGNs, if the gas close to the SMBH joins an accretion disc
with a standard radiative efficiency of $\sim 0.1$ (e.g., Loewenstein
et al. 2001; Fabbiano et al. 2003). Since these luminosities are not
observed, the options left are that the radiative efficiency is orders
of magnitude lower, as in an advection dominated accretion flow (ADAF)
and its variants (e.g., Narayan 2002), or that accretion is unsteady
(Binney \& Tabor 1995, Ciotti \& Ostriker 2001). In addition, nuclear
jets can carry away a large fraction of the estimated accretion power,
a possibility that has been found very reasonable for M87 (Di Matteo
et al. 2003), IC4296 (Pellegrini et al.  2003b) and IC1459 (Fabbiano
et al. 2003).  NGC~821 represents another promising case to test the
hypotheses above.

NGC~821 is an E6 galaxy with possible disky isophotes and centrally
peaked (power-law) surface brightness (Ravindranath et al 2001). At a
distance of 23~Mpc (e.g., de Zeeuw et al. 2002), the ACIS resolution
at the aim point corresponds to 55~pc. Prior to our {\it Chandra}
observation, NGC~821 had been observed, but not detected, in X-rays
with {\it ROSAT} ($ < 5 \times 10^{40} \rm erg~s^{-1}$; Beuing et
al. 1999). This limit is $\sim 10^5$ times below the Eddington luminosity
of the nucleus, based on the SMBH mass of $2.8-5.8 \times 10^7
M_{\odot}$, which was measured from
HST (STIS) spectra of the nuclear region coupled to dynamical
modeling (Gebhardt et al. 2003).  NGC~821 was observed with {\it Chandra} as
part of a mini-sample of extremely faint SMBHs extracted from the list
of Ho (2002), which will be the subject of a follow-up paper. As
reported in Ho (2002) and Ho et al. (2003), NGC~821 has not been
detected in radio continuum nor in optical emission lines (H$\alpha$
and H$\beta$), and is a good example of quiescent SMBH. Here we report
the results of the {\it Chandra} ACIS observation of NGC~821, that has led
to the discovery of an S-shaped feature, suggestive of either a weak
two-sided X-ray nuclear jet, or of hot shocked gas.

\section{Observations and Data Analysis}

NGC~821 was observed with {\em Chandra} (Weisskopf et al 2000) ACIS-S
(PI: Fabbiano) on November 26, 2002 (ObsID: 4408) and on December 1,
2002 (ObsID: 4006) for a total exposure time of 38~ks.
Table~\ref{obslog} is a summary of the relevant properties of
NGC~821 and of the observing log.

\begin{deluxetable}{lccccccccc}
\tabletypesize{\scriptsize}
\tablecaption{NGC~821: Properties$^a$ and {\it Chandra} ACIS-S Observation Log 
\label{obslog}}
\tablewidth{0pt}
\tablehead{
\colhead{M$^0_{B_T}$} &
\colhead{D} &
\colhead{Diam.(')} &
\colhead{N$_H$} &
\colhead{L$_X$(erg~s$^{-1}$)}&
\colhead{$M_{\bullet}$}&
\colhead{0".5}&
\colhead{ObsID }&
\colhead{Date}&
\colhead{T$_{exp.}$}
\\
\colhead{(mag)} &
\colhead{(Mpc)} &
\colhead{$\sigma_o$(km~s$^{-1}$)} &
\colhead{(cm$^{-2}$)} &
\colhead{$L_{H\alpha}$(erg~s$^{-1}$)}&
\colhead{($10^7 M_{\odot}$)}&
\colhead{(pc)}&
\colhead{}&
\colhead{}&
\colhead{(ks)}
}
\startdata
-20.71&23&2.6&6.2$\times 10^{20}$&$<$5$\times 10^{40~b}$&2.8-5.8$^{~c}$&55&4408&Nov. 
26, 
2002& 24.6\\
\nodata &\nodata &209$^d$ &\nodata &$< 1.3 \times 10^{38}$$^e$ &\nodata &\nodata 
&4006& Dec. 1, 2002&13.3\\
\enddata

$^a$ Unless otherwise noted, the galaxy properties are as listed in NED (NASA / IPAC 
Extragalactic Database);
$^b$ Beuing et al 1999; $^c$ Gebhardt et al. 2003, rescaled for the distance
adopted here; $^d$ Prugniel \& Simien 1996;
$^e$ Ho et al. 2003

\end{deluxetable}

No significant background flares were observed in these data, so no
further screening was necessary.  A time-dependent gain correction
({\footnotesize URL: http://cxc.harvard.edu/contrib/alexey/tgain/tgain.html}) was 
applied
to the Standard Data Processing Level 2 event files, before further
analysis.  The data were then analyzed using the CIAO v3.0.1 software (CALDB 2.23)
and XSPEC v11.2.0 for the spectral fits.

\subsection{X-ray image}

The two observations were merged, taking into account the relative
aspect solution of the two data sets, with the CIAO task
$merge\_all$. From the resulting dataset, images were extracted in
three spectral bands (Red = 0.5 - 1~keV; Green = 1 - 2~keV; Blue = 2 -
4~keV), covering the spectral range in which most source counts are
detected.  The images were then smoothed using {\it csmooth} with scales
ranging from 1 to 20 pixels (0.5$^{\prime\prime}$- 10$^{\prime\prime}$).  Fig.~1 
shows the
resulting `true-color' X-ray image of a 3' field centered on
NGC~821. Given the small region considered, no exposure correction was
needed. The ellipse indicates the $D_{25}$ isophote from RC3. Most of
the X-ray emission is concentrated in the central regions of the
galaxy, and it includes a diffuse component, a few point-like
sources, and a brighter central/nuclear region.

A higher resolution `true-color' image of the central region of
NGC~821 is shown in fig.~2a, where the data are displayed at the
original observed resolution, without smoothing. This figure shows
clearly a north-south elongated, hard central feature. Within the
{\it Chandra} aspect uncertainties ($<0.5^{\prime\prime}$), this feature
is centered on the nucleus of NGC~821, at RA=$02^h 08^m 21.14^s$,
Dec=$+10^o 59^\prime 41.7^{\prime\prime}$ (J2000, with uncertainty
of $1.25^{\prime\prime}$; from the 2MASS survey, as reported in NED). While a
point-like source may be superposed on the southern tip of this
elongated feature, the general form is suggestive of a two-sided bent
jet, or S-shaped filament centered on the nucleus. 
This feature is approximately 8.5'' long, corresponding to $\sim
1$~kpc at the distance of NGC~821.

\subsection{Point source detection and spatial analysis of the nuclear
feature}

We ran the CIAO {\em wavdetect} tool on the full-band ACIS image, with
scales between 2 and 4 pixels (1-2"), and detected 11 sources (above
3~$\sigma$) in the area shown in fig.~1.  We assume a power-law
spectrum with $\Gamma = 1.8$ and Galactic $N_H =
6.4\times10^{20}$~cm$^{-2}$ (Stark et al 1992), consistent with the
emission of low-mass X-ray binaries (LMXBs) detected in early-type
galaxies (see Kim \& Fabbiano 2003). With this spectrum, the detected
 sources have 0.3-10~keV
`emitted' fluxes in the $1.4\times10^{-15}-1.4\times10^{-14}$ erg
cm$^{-2}$ s$^{-1}$ range, corresponding to luminosities of
$1.2\times10^{38}-1.2\times10^{39}$ erg s$^{-1}$, if they indeed
belong to NGC~821. The faintest sources we can detect in our exposure of
NGC~821 are at the upper end of the luminosity distribution of the 
populations of LMXBs detected in other elliptical galaxies with {\it Chandra}
(see e.g., Kim \& Fabbiano 2004).
Based on these results, we expect that most LMXBs in NGC~821 
will have $L_X \leq 1 \times 10^{38} \rm erg~s^{-1}$
and therefore will be undetectable. These LMXBs, however, will contribute to 
the unresolved `diffuse' galaxian emission.

In the central region shown in fig.~2a, there are four sources: the
isolated point-like source at the north-east of the central complex
(source NE, with 0.3-10~keV $L_X = 4.7\times10^{38}$ erg s$^{-1}$),
and three sources in the central elongated emission region; these are
identified by ellipses in fig.~2b, and named S1, S2 and S3, going from
north to south.  Further analysis (see below) demonstrates that these three
central emission regions are not point-like and therefore cannot be
explained with the serendipitous positioning of three luminous
galaxian LMXBs in NGC~821. No discernible features can be seen in the optical
image of NGC~821 at the positions of these three sources. Using the Deep Survey 
source counts in the 0.5-2~keV band (Rosati et al 2002), the number of expected 
sources at fluxes of $\geq 1.4\times10^{-15}$~erg
cm$^{-2}$ s$^{-1}$ is $\leq 8 \times 10^{-3}$, so that chance detection of 
background sources is very unlikely. We cannot exclude a peculiar 
clustering of very luminous galaxian X-ray binaries, and future deeper
data will be needed to explore this point further.

To establish the spatial properties of sources S1, S2 and S3, we have
compared the spatial distribution of counts from each of them with
that of the on-axis image of the quasar GB~1508+5714 (Siemiginowska et
al 2003a). GB~1508+5714 can be used as a good representation of the
{\it Chandra} ACIS-S PSF for our analysis, since the sources we are
interested in are also at the aim point and similarly hard. With a
count rate of $\sim 0.05 \rm~count~s^{-1}$, the image of GB~1508+5714 (ObsID 2241)
is not affected by pile-up, and contains 5,300 counts within 2'' of the
centroid of the count distribution. From this image, we determine the
ratio of counts within the 1"-2" annulus to those in the central 1"
radius circle to be {\it Ratio}(PSF)$ = 0.043 \pm0.001~(1\sigma)$. The isolated 
NE source in
this central field yields {\it Ratio}(NE)$ = 0.057 \pm 0.054$ (from a total
of 36 source counts), entirely consistent with that of our reference
quasar, confirming that GB~1508+5714 gives a good representation of the PSF. 
Instead, the
analogous ratios for the background-subtracted counts from S1, S2 and
S3 demonstrate that the emission is extended in all cases.
Using the {\it wavdetect} centroids, we obtain {\it Ratio}(S1)$ =  0.81 \pm 0.26$,
{\it Ratio}(S2)$ =  0.94 \pm 0.27$, and {\it Ratio}(S3)$ =  0.29 \pm 0.10$. The
total number of source counts in the three cases are 56, 67, and 71,
respectively, larger than for source NE.  This comparison demonstrates that the 
spatial
distributions of the source counts from S1 and S2 are definitely not
consistent with the PSF. Although the spatial count distribution of S3
is more peaked and could contain a point-like component, diffuse
emission is also present. This analysis reinforces our suggestion that the central
emission feature is intrinsically elongated.

Given that the central emission (S2) is not consistent with a
point-source, we can only estimate a 3~$\sigma$ upper limit on the
luminosity of a nuclear point-like AGN.  We used a $1" \times 1"$ ($2
\times 2$ pixels) sliding cell over the entire area covered by the
S-shaped feature, and assumed as the background level the maximum value detected
in the sliding cell. This conservative approach yields a  0.3-10~keV 
$L_X < 4.2 \times 10^{38} \rm erg~s^{-1}$ for a power-law spectrum
with $\Gamma =1.8$ and line of sight Galactic $N_H$. This limit indicates 
that a central point-like AGN would have a luminosity not exceeding 
that of normal LMXBs. 

\subsection{Spectral analysis}

We extracted spectral counts both from the S-shaped feature
(regions in fig.~2b) and the surrounding diffuse emission. The latter was taken
from a 20'' circular region centered on the nucleus, excluding the S-shaped
feature and the
NE source.  The field background was extracted from a source-free 50''
radius circular area 1.8' from the nucleus of NGC~821.  Spectral
counts were extracted separately for the two observations using the
CIAO script {\em acisspec}, which extracts a spectrum for both the
source and the background region and creates weighted response
matrices and ancillary response files (ARF). The ARFs were corrected
for the time-dependent degradation in the ACIS quantum efficiency, using
the CIAO script {\em apply\_acisabs}.  The source spectra from the two
observations were then coadded (using the FTOOL {\em mathpha}) to
optimize the signal-to-noise ratio of the data.  We used the FTOOLS
{\em addarf} and {\em addrmf} to combine the responses with their
appropriate weights.

%


Hardness ratios (HR1=M-S/M+S; HR2=H-M/H+M; where S=0.5-1 keV, M=1-2
keV, and H=2-4 keV) are plotted (with 1~$\sigma$ statistical errors)
in fig.~3, and compared with power-law and Raymond-Smith emission
models. The hardness ratios of the S1, S2, S3 regions are all
consistent with hard emission, either a power-law spectrum with $\Gamma$ between 
$\sim1$ and
$\sim2.2$, or thermal emission with $kT \sim 2-20$~keV. 
The diffuse X-ray emission of an elliptical galaxy is the
combination of the soft emission of the hot ISM and of the hard emission of
the population
of LMXBs below our detection threshold (see e.g, Kim \& Fabbiano
2003). In NGC~821 the hardness ratios of the diffuse emission 
are hard, suggesting a dominant LMXB component and little hot ISM.

With XSPEC, we fitted the data in the 0.3-10~keV energy range,
rebinned to have at least 15 counts in each energy bin. For the S-shaped 
feature we adopted an absorbed power-law model (XSPEC model: {\em
wabs(wabs(pow))}), with $N_H$ consisting of both a Galactic and an
intrinsic component. 
The results of the fits are listed in Table~\ref{fitres}, with 90\%
errors on one significant parameter. The  S-shaped  emission is well fitted
with a power-law spectrum with $\Gamma \sim 1.8$, typical of AGN
spectra, although the uncertainties are large.  As discussed in
\S~2.2, there could be a possibly unrelated point-like source at the
southern tip of the S-shaped feature. Given the very few counts detected in our
observations, we were unable to perform a meaningful spectral
analysis, eliminating this source.  However, the X-ray hardness ratios
of fig.~3 show that the X-ray colors for the three regions S1, S2, S3 are
remarkably similar, and consistent with the power-law model derived
from the minimum-$\chi^2$ fit.

\begin{deluxetable}{ccccccc}
\tabletypesize{\small}
\tablecaption{Results of Spectral Fits \label{fitres}}
\tablewidth{0pt}
\tablehead{
\colhead{} &
\colhead{\begin{tabular}{c}
Net Counts\\
(0.3-10 keV)
\end{tabular}} &
\colhead{$\chi^2/dof$} &
\colhead{\begin{tabular}{c}
$N_H$\\
($\times10^{21}$ cm$^{-2}$)
\end{tabular}} &
\colhead{\begin{tabular}{c}
kT\\
(keV)
\end{tabular}} &
\colhead{$\Gamma$ } &
\colhead{\begin{tabular}{c}
$L_X (0.3-10~keV)$\\
(erg~s$^{-1}$)
\end{tabular}}  
}
\startdata
S-shaped & 141 & 6.7/5 & $1.41_{-1.41}^{+1.84}$ & \nodata & $1.82_{-0.56}^{+0.71}$ & 
$2.6\pm0.5\times10^{39}$\\
... & ... & 6.5/5 & $ 0.75_{-0.75}^{+2.34}$ & $5.14_{-2.99}^{+50.00}$& ...& 
$1.9_{-0.4}^{+1.4} \times 10^{39}$\\
Diffuse & 174 & 6.5/10 & $<9.00$ & $0.46_{-0.25}^{+0.33}$ & $1.27_{-0.68}^{+1.13}$ & 
$3.5\pm0.6\times10^{39}$\\
Thermal &\nodata &\nodata&\nodata&\nodata&\nodata& $3.4\pm0.1\times10^{38}$\enddata

\end{deluxetable}

For the diffuse emission, following e.g. Kim \&
Fabbiano (2003), an optically thin thermal component was added to the
power-law model (XSPEC model: {\em wabs(wabs(apec+pow))}).  The metal
abundance of the thermal component was fixed to 0.3 times the solar
value (keeping the elemental solar ratios of Anders \& Grevesse 1989).
The spectrum of the diffuse galaxian emission (see Table~\ref{fitres})
is consistent with a hot gas with
a temperature $kT\sim 0.5$~keV, or cooler, typical of X-ray-faint
early-type galaxy halos (e.g. Pellegrini \& Fabbiano 1994, Irwin \&
Sarazin 1998). The power-law component, although ill-defined, is
needed to obtain an acceptable fit for the diffuse emission, as clearly
suggested by fig.~3; this is
in agreement with the presence of an unresolved LMXB population. The
$\chi^2$ for a simple thermal model ($APEC$; Smith et al 2001) is 25.8
with 12 degrees of freedom. Comparing this with the two-component
$\chi^2$ by means of an F-test, we obtain a chance probability of
$10^{-3}$. Fitting the diffuse emission with a single power-law model,
also results in a worse $\chi^2$, but in this case the F-test
probability is only 12\%. While the two-component model is plausible
for the diffuse emission, the rather low signal-to-noise ratio of our
data does not allow a stronger discrimination among
models. In the following discussion we will use the results of the
two-component model fit; however, future deeper observation of NGC~821 
are essential to firmly constrain the characteristics of the hot ISM
in this galaxy. 

Table~\ref{fitres} also lists the best-fit unabsorbed
luminosities for the S-shaped feature, the total diffuse emission and the thermal 
component of the diffuse emission. The latter is only 10\% of the total diffuse 
emission, indicating that NGC~821 is singularly devoided of 
hot ISM. Strictly speaking, we have a (90\%) upper limit of 3.5$\times10^{38} \rm 
erg~s^{-1}$
on the luminosity of the gaseous component. The uncertainties on these
luminosities reflect the uncertainties on the spectral parameters.
From the emission measure of the soft thermal component 
we estimate an `indicative' electron density $n = 4.1_{-1.5}^{+10.9} \times
10^{-3} \rm cm^{-3}$. While future deeper data are needed for a definite measure,
the above estimate is useful for gaining first-cut insights on the nature of
the S-shape feature, and in this spirit we will use it in the following
discussion.

\section{Discussion}

NGC~821 was observed with {\it Chandra} because the existing {\it ROSAT} upper
limit on the X-ray luminosity of its nuclear SMBH implied extremely
sub-Eddington nuclear emission. Our observations fail to detect a
point-like source at the nucleus, down to a $3 \sigma$ limit of
$L_X(0.3-10~keV) < 4.2 \times 10^{38} \rm erg~s^{-1}$, $\sim 100$
times fainter than the ROSAT limit (see Table~1). The nucleus is not
detected in the FIR or in H$_2$ (Georgakaki et al. 2001) arguing
against a strongly obscured AGN (and against a nuclear
starburst). There is also no sign of strong intrinsic absorbing column
in the X-rays (see Table~2). The general AGN `quiescent state' is also
supported by the lack of radio continuum and optical line emission (Ho
2002; Ho et al. 2003).  We detect instead an elongated, possibly bent,
emission feature, strongly suggestive of a two-sided X-ray jet or
S-shaped filament.  This feature, extending over $\sim1$~kpc, has a
hard spectrum consistent with a power-law with best-fit $\Gamma \sim
1.8^{+0.7}_{-0.6}$ ($\alpha$=0.8) or, if thermal, $kT>2$~keV.  The
X-ray luminosity of this feature is $\sim 1.9-2.6 \times 10^{39} \rm
ergs~s^{-1}$, corresponding to $\sim 5 \times 10^{-7}$ of the
Eddington luminosity of the SMBH.

The nature of this intriguing elongated feature is investigated
below, considering the possibilities of a two-sided jet (\S~3.1)
and of thermal emission from gas heated by some form of energy deposition 
resulting from  nuclear activity (\S~3.2 and 3.3). 

\subsection{Is a jet directly emitting the hard X-rays?}

We discuss here the possibility that the S-shaped, hard emission
centered on the nucleus of NGC~821 is a two-sided nuclear jet, as in
radio galaxies.   The spectral power-law slope of
this emission has large uncertainties (Table~2), but is consistent
with the X-ray spectra of other jets (Siemiginowska et
al. 2003b; Sambruna et al 2004). However, unlike other extragalactic X-ray jets seen in luminous
AGNs, where $L_X(jet)/L_X(AGN)\sim $ 1--15\% (e.g., Schwartz et
al. 2003), this `jet' has no associated core X-ray source, implying
$L_X (jet) / L_X (AGN) > 6$. This lack of core emission also differs
from other X-ray weak AGNs, where  only a point X-ray source has been seen, 
typically in nuclei with radio detections, and these are usually modelled as
Comptonized jets (Baganoff et al. 2001, Markoff et al. 2001, Fabbiano et al 2003). 
An exception is M~87, with its core-jet structure (Di Matteo, Carilli \& Fabian 2001).
The NGC~821 `jet' could be similar to the somewhat steeper spectrum
($\Gamma\sim$2.3) M87 jet (Wilson \& Yang 2002), where the nuclear
point-like AGN is fainter compared with the jet ($L_X (jet) /
L_X (AGN)\sim 2$).  The 0.5~kpc
(one-sided) scale of the NGC~821 `jet' is also similar to the
$\sim$1.5~kpc of the M~87 jet, while most radio/X-ray jets extend over
much larger distances, up to 300~kpc (Siemiginowska et al. 2002, 2003b). The
NGC~821 `jet' is two-sided, putting a limit on relativistic beaming
and suggesting only a relatively slow expansion ($v < 0.3c$; see
e.g. Leahy 1991).  The typical expansion velocity of the M87 jet is
0.5$c$, though small features can reach 4-6$c$ (Biretta et al. 1995,
1999).

If a jet is present, we can make some simple estimates of the activity
timescale and of the energetics involved.  First, based on the above
limit on the expansion velocity ($v<0.3c$), we can relate the size of
the jet to the activity timescale, finding that the jet may have been
propagating for a few thousand years, assuming free unimpeded
propagation.  However, it is more likely that the jet is disrupted by
interaction with the surrounding hot ISM ($n = 4.1_{-1.5}^{+10.9}
\times 10^{-3} \rm cm^{-3}$) as, for example, may be occuring in
NGC~1316, which has a similar ISM density (0.01~cm$^{-3}$ in the jet
disruption region; Kim, Fabbiano \& Mackie 1998).  Theoretical
simulations (De Young 1993) of jet propagation into a homogeneous ISM
indicate that hot ISM with densities of 0.01 particles~cm$^{-3}$,
similar to those of NGC~821, can significantly slow down a jet 10$^5$
times more powerful than the one we may have in NGC~821.

Considering that the jet is likely to interact with the surrounding
hot ISM, its size may be indicative of equilibrium between the ram
pressure of the expanding jet and the thermal pressure of the hot ISM.
We estimate the thermal pressure of the ISM from the diffuse X-ray
emission to be $\sim 1.5 \times 10^{-11}$ dyn~cm$^{-2}$ ($P_{th}$=$2
n_e kT$). Equating this pressure to the ram pressure $P_{ram} =
\rho_{ISM} v^2$, where $\rho_{ISM}$ is the mass density of the ISM, we
derive a jet expansion velocity $v$ of $0.043c$, consistent with the
upper limit mentioned earlier, and corresponding to a jet expansion
(or activity) timescale of 3.8$ \times 10^4$ years.  The jet kinetic
energy can be estimated from $E_{kin}=0.5 P_{ram} V$, where $V$ is the
volume occupied by the jet. To calculate $V$, we assumed a jet
thickness of 10~pc and a length of 1~kpc, obtaining $V=$2.7$\times
10^{60}$ cm$^{3}$. The kinetic energy input to this volume is then
4.1$\times 10^{49}$~erg, and by dividing this $E_{kin}$ by the jet
expansion time we derive a kinetic luminosity of $1.7\times 10^{37}$
erg s$^{-1}$. This is $\sim 100$ times smaller than the X-ray
luminosity. However, this may be a lower limit if the jet is
``overpressured'', as in M87 where the radio cavities in the cluster
gas suggest overpressurization by a factor $\sim 3$ with respect to
the cluster gas (Reynolds 1997; Reynolds et al 1996). However, M87 has a
jet kinetic luminosity of $\sim 10^{43}$~erg~s$^{-1}$, and a radiative
power of $\sim 10^{42}$~erg~s$^{-1}$, far more powerful jet than in
NGC~821.

What would be the emission mechanism responsible for the S-shaped
emission of NGC~821, in the jet hypothesis? Synchrotron emission is plausible: 
$\alpha_{radio-X} \leq 0.7$ (estimated from the radio
flux limit of 0.5~mJy at 5~GHz and the X-ray flux at 1~keV);  this
$\alpha_{radio-X}$ is consistent with the X-ray slope, and is typical
of the synchrotron slope observed in radio lobes (Peterson 1997). The
usual problem of short synchrotron lifetime for X-ray producing
electrons (Feigelson et al. 1981), imply local particle
re-acceleration 0.5~kpc from the nucleus. A radio detection not far
below the current limit is expected if synchrotron radiation produces
the S-shaped emission at the center of NGC~821. Also in this respect, the
NGC~821 emission could be consistent with the M~87 jet:
Wilson \& Yang find $\alpha_{radio-X}\sim 0.9$ for the knots of the
M87 jet, and they prefer a synchrotron origin for the X-ray emission
of these knots.  

However $\alpha_{radio-X} \leq 0.7$ is also consistent with the values
of $\alpha_{radio-X}$=0.7-0.8 reported for knots in powerful jets by
Sambruna et al. (2004) who favor an Inverse Compton origin for the
X-ray flux for most of the jet knots, based on the X-ray fluxes lying
above an extrapolation of the radio-optical slope ($\alpha_{RO} >
\alpha_{OX}$).  So the value of $\alpha_{radio-X} \leq 0.7$ may be
merely a coincidence, and is in any case only a limit, so it is
worthwhile to consider an Inverse Compton origin. The seed photons for
Comptonization could be either from the synchrotron photons within the
jet (the `self-Compton' case) or could be the external to the jet photon field.
In the self-Compton process the ratio of the synchrotron to Inverse
Compton luminosities is given by the ratio of energy densities
of the magnetic to the synchrotron radiation field. 
In equipartition both luminosities are of the same order, and since
the observed X-ray luminosity is a factor of at least 10$^3$ higher than
the radio upper limit, the self-Compton case is excluded.

External Comptonization is also ruled out: Felten \& Morrison (1966,
eq.47) showed that $I_S/I_C = U_B/U_{ph}(\nu_C/\nu_S)^{(3-m)/2}$. Here
$I_S$ and $I_C$ are intensities of synchrotron and inverse Compton
emission, $U_B$ is the energy density of magnetic field, $U_{ph}$ the
energy density of the external photon field, $\nu_C/\nu_S$ is the
ratio between the frequency of the Compton scattered photon and the
frequency of the synchrotron photon, $m$ is the power law index of the
electron distribution, which is linked to the spectral index,
$\alpha_s$ of the synchrotron emission so $m=1-2\alpha_s$.  We measure
$I_C$ as the X-ray luminosity, and $U_{ph}$ from the starlight from
the central cusp of NGC~821 (Gebhardt et al. 2003, rescaling their V-band 
luminosity density for the distance in Table~1). 
The maximum synchrotron luminosity, $I_S$ comes from the
0.5~mJy flux limit at 5~GHz, integrating over the 10$^7-10^9$~GHz
range with a slope of 0.7, and it is $<$4.5$\times 10^{36}$erg~s$^{-1}$.
$U_B$ can be derived from the radio
upper limit assuming equipartition of the magnetic field and the
electrons.  Because the jet expansion is not relativistic we can
assume the bulk jet Lorentz factor $\Gamma \sim 1$, simplifying
comparison of the radiation fields. Assuming equipartition (Burbidge
1959) between the $U_B$ and the electron energy density, gives
the minimum magnetic field (Krolik, 1999, eq. 9.21), which can be expressed as:
$B_{min}=4819.84 (L_0/V)^{2/7} \nu_0^{-1/7} ({\nu_l \over
\nu_0}))^{(1-2\alpha_s)/7}$~G, where in our case $\nu_0 = 5$~GHz, 
$\nu_l$ is the lowest frequency of the synchrotron emission,
L$_0$ is the monochromatic luminosity at 5~GHz, and $V$ is the emission volume.
We obtain: $B_{min}$=~2.5$\times10^{-5}$G, and $U_B= \sim 2.4\times
10^{-11}$erg~cm$^{-3}$.

Considering the radial dependence of the optical photon field,
we obtain a maximum
predicted Inverse Compton emission due to scattering of the starlight
ranging from $3.7 \times 10^{36} \rm erg~s^{-1}$
at 1~pc galactocentric radius, to 
$7.9 \times 10^{35}  \rm erg~s^{-1}$ at 500~pc (the maximum jet extension). 
We conclude that 
Inverse Compton radiation would be a small contribution to the X-ray emission, suggesting 
that synchrotron may be the dominant emission mechanism if the S-shape feature is
indeed a jet.

\subsection{Is hot gas responsible for the hard emission?}

We examine here the possibility that the origin of the hard S-shaped
emission is thermal. We consider two scenarios suggested by analogies with
other galaxies: thermal emission from the jet-ISM interaction, as in 
NGC~1052 (Kadler et al 2004), and shocks in the ISM, resulting from an
outburst of nuclear activity, as suggested in the case of NGC~4636 (Jones
et al 2002).

An elongated feature in the central galaxy region, of temperature
$kT\sim$0.5~keV, was found with {\em Chandra} in NGC~1052, the
prototype elliptical galaxy LINER (also at a distance of 22.6~Mpc, Knapp et
al. 1978). This feature has 10 times the flux of the S-shaped emission  in NGC~821 
($\sim$3.5$\times$10$^{-13}$erg~cm$^{-2}$s$^{-1}$), and
about double the linear size (Kadler et al. 2004). In NGC~1052 the
extended emission has radio and optical counterparts, and may be more
fan shaped than linear, though the photon statistics are limited
in this short (2.3~ksec) ACIS observation. Kadler et al. suggest
shock heating of gas, converting some of the kinetic
energy of the observed radio jet into X-ray emission.  
In the case of NGC~821, if a jet is present, it could similarly deposit
energy in the surrounding ISM, causing it to produce the hard thermal
emission. However, if its kinetic power is of the order of that
estimated in Sect. 3.1 ($\sim 10^{37}$ erg s$^{-1}$), this mechanism
cannot explain the luminosity of the hard thermal emission measured
with ACIS ($L_X\sim 1.9\times 10^{39}$ erg s$^{-1}$, Table 2).  
Unless a jet is present with orders of magnitude larger kinetic power, other
sources of energy deposition are needed.

An alternative analogy is offered by the hot `arms' of NGC~4636 (Jones et al 2002), 
a giant elliptical in Virgo with no reported nuclear activity or jets. These arms, 
having a larger spatial scale (8~kpc) and a lower
temperature ($\sim 0.5 - 0.7$~keV) than the S-shaped feature of
NGC~821, are two symmetric features crossing the
galaxy center, discovered in the {\em Chandra} ACIS data of NGC~4636. The 
NGC~4636 arms are accompanied by a temperature increase with respect to the 
surrounding hot ISM,  which led to the suggestion by Jones et al of shock 
heating of the ISM caused by a nuclear outburst. We will discuss this process
at the end of Sect. 3.3 below, in the unsteady accretion hypothesis.
As discussed in \S~2.3, the S-shaped feature of NGC~821 could similarly be hotter than 
its surroundings. Assuming a temperature of kT=3~keV for this feature, close to the
lower limit suggested by our spectral fit (Table~2), we 
obtain a density $ n = 8.59_{-1.07}^{+1.06} ~\rm cm^{-3}$ (errors at 90\%).
Using the best-fit value (and a temperature of 3~keV), we obtain 
a thermal pressure exceeding by a factor of $\sim 14$ that quoted in \S 3.1 for the
surrounding hot ISM, suggesting a non-equilibrium situation, if only thermal
pressures are involved. However, given the uncertainties in both $T$ and $n$, the 
thermal pressures of both S-shaped feature and surrounding ISM could be similar.
It is clear that deeper {\it Chandra} observations are needed to better constrain the
energetics of this feature.


\subsection{Is accretion present?}

Whether it is a jet or hot shocked gas, the S-shaped feature of
NGC~821 requires a considerable energy input, an obvious source of
which is accretion onto the nuclear SMBH.  Taking at face value the
indication of the two-component fit of the circum-nuclear diffuse
emission, which is consistent with the presence of a $kT\sim 0.5$ keV
thermal component, we can estimate the nuclear accretion rate, in the
steady spherical accretion scenario of Bondi (1952). This estimate is
rough, because the gravitational capture radius (which depends on the
gas temperature and on the mass of the SMBH; see the textbook by Frank,
King \& Raine 2002) is $r_{acc}\sim 3-23$ pc in our case, smaller than
the physical resolution of the image (55~pc, Table~1). In this
estimate of $r_{acc}$ we have allowed for the uncertainties in both
the SMBH mass and $kT$. Based on the circum-nuclear gas
temperature and density, that we estimate from the emission
measure of the diffuse thermal component to be $n = 4.1_{-1.5}^{+10.9} \times
10^{-3} \rm cm^{-3}$, the Bondi mass accretion
rate (also following Frank, King \& Raine 2002) is $\dot M
_{acc}~=1.1~\times~10^{-7} - 2.0~\times~10^{-5}~\rm M_{\odot}~\rm
yr^{-1}$, including again all the uncertainties in the SMBH mass, $kT$ and
$n$. The corresponding luminosity is $L_{acc}~=~6.2 \times 10^{38} -
1.1 \times 10^{41}\, \rm ergs~s^{-1}$, with the customary assumption of
10\% accretion efficiency.  
The luminosity of the S-shaped feature is within this
range, therefore in principle it could be explained by Bondi accretion
of the hot ISM. This conclusion, however, is most certainly not correct, because 
comparison with independent optical data, discussed below, shows that our uncertain 
X-ray data may have led to underestimating the circum-nuclear gas density and 
therefore   $\dot M_{acc}$.

Since the hot ISM is the thermalized integrated result of the stellar
mass losses, as a minimum one expects the total stellar mass loss rate
$\dot M_{\star}$ within $r_{acc}$ to be accreted (there may also be gas
inflowing from outside $r_{acc}$).  $\dot M_{\star}$ can easily be
obtained from the luminosity density profile recovered from HST data
for the central galaxy region (Gebhardt et al. 2003). Using
a conversion factor from luminosity to mass loss rate for an old
stellar population at the present epoch (e.g., Ciotti et al. 1991),
this leads to $\dot M_{\star} = 9.8\times 10
^{-6}$ and $2.6\times 10^{-4}$~M$_{\odot}$yr$^{-1}$,  for the two
extreme values of $r_{acc}$. These $\dot M_{\star}$ values are
respectively a factor of $\sim$90 and $\sim$13 larger than the $\dot
M_{acc}$ values derived above for the same $r_{acc}$. Given that this
estimate of $\dot M_{\star}$ is quite robust, we must
conclude that either the derivation of $\dot M_{acc}$ above is
inaccurate, or the gas is not steadily inflowing within $r_{acc}$.
The former possibility cannot be excluded with the present data, since in our 
calculation of  $\dot M_{acc}$ we used a density $n$ value that is an
average measured over a region extending much farther out than $r_{acc}$;
$n(r_{acc})$ is likely to be significantly higher than this average 
(e.g., a factor of $\sim 30 $ times higher, for a
$n\propto r^{-0.9}$ profile). 

Assuming that it is just the stellar mass loss rate
within $r_{acc}$ that is steadily accreted, accretion luminosities
$>20$ times larger than the observed $L_X$ of the hard emission are
recovered (from $L_{acc}=0.1\dot M_{\star}c^2$).  Then, the
"solutions" proposed for the other X-ray faint nearby nuclei must be
revisited for NGC~821. These are (a) accretion occurs but with low
radiative efficiency (e.g., Narayan 2002); (b) accretion sustains a
jet (this can be coupled again to a low radiative efficiency), whose
total power is of the order of $L_{acc}$, as in the modeling for
IC~1459 (Fabbiano et al. 2003) and M87 (Di Matteo et al. 2003); (c)
accretion is unsteady and therefore the hot ISM in the nuclear region
needs not be inflowing (Siemiginowska et al 1996; Janiuk et al 2004). 
In this case the feedback from the central
SMBH can be either radiative (Ciotti \& Ostriker 2001) or mechanical
(Binney \& Tabor 1995, Omma et al. 2004) and make accretion undergo
activity cycles: while active, the central engine heats the
surrounding ISM, so that radiative cooling -- and accretion -- are
offset; then the central engine turns off, until the ISM starts
cooling again and accretion resumes.  NGC~821 may be in a stage of
such a cycle when a nuclear outburst has recently occurred. Note
that the accretion luminosity that is radiatively absorbed by the
ISM during an outburst (Ciotti \& Ostriker 2001) largely exceeds 
the hard thermal emission observed at the center of NGC~821.

The presence of the central S-shaped feature that is
{\em hotter} than the surrounding gas is uncontroversial evidence that
central heating is at work, and therefore some type of feedback from
the SMBH is occurring. The unsteady scenario seems promising to fit
adequately the case of NGC821, also because this galaxy is clearly hot gas poor,
as if recently swept by an outburst-driven wind.  
From our ACIS-S data we estimate un upper limit of $\sim 4
\times 10^6 M_{\odot}$ on the amount of hot ISM, many orders of
magnitude smaller than for hot gas rich ellipticals (see Fabbiano 1989).

\section{Conclusions}

We have shown that NGC~821, which is used as template of a quiescent
and normal `old elliptical' (see, e.g. Ho et al 2003) shows clear signs of
energy deposition by the nucleus: 
an elongated, bent, kiloparsec-size S-shaped feature, centered (within the $\sim
1.5^{\prime\prime}$ errors) on the nucleus. This feature has a (0.3-10 keV) 
$L_X \sim 1.9-2.6 \times 10^{39} \rm
ergs~s^{-1}$, and a hard spectrum consistent with a power-law with $\Gamma
\sim 1.8$, or thermal emission with $kT > 2$~keV. It may be embedded in a faint
hot ISM with $kT \sim 0.5$~keV.

This feature could represent the faintest yet reported extragalactic
jet.  We can exclude Inverse Compton (either of radio photons or of the central
stellar photon field) as an explanation of the X-ray emission. If this feature
is indeed a non-thermal jet, this leaves synchrotron as a possible explaination.
The $\alpha_{radio-X} \leq 0.7$ would be consistent with this hypothesis.
In the jet scenario, we constrain the activity timescale to be $\sim 4 \times 
10^4$~years. We also estimate that in this jet most of the power may be in
X-ray radiation, rather than kinetic energy.

Alternatively, the emission could be thermal. It is unlikely that the
detected luminosity could result from interaction of a jet with the surrounding
ISM in this galaxy (as may occur e.g., in NGC~1052, Kadler et al 2004) because the 
radio upper limits imply a total jet kinetic power orders of magnitude below the 
detected X-ray luminosity. A more likely scenario is that of S-shaped shocks of the 
hot ISM, driven by an outburst of nuclear activity, as suggested in the case of 
NGC~4636 (Jones et al 2002). 
The characteristics of this feature (luminosity and hard spectrum) suggest that it is 
driven by nuclear activity. Accretion from stellar mass loss rate in the 
circum-nuclear region would produce a luminosity well in excess of the detected one,
suggesting either low radiative efficiency (e.g., Narayan 2002) or unsteady accretion
(e.g., Binney \& Tabor 1995; Siemiginowska et al 1996;
Ciotti \& Ostriker 2001). The second scenario is 
supported by the very small amount of hot ISM
that may be present in NGC~821. 
It remains unexplored by the current modeling of outbursts through
hydrodynamical simulations whether and how an S-shaped structure can be
created.

It is only in the
X-ray band, thanks to the {\em Chandra} spatial resolution, that signs of
nuclear activity are found in NGC~821. Either jet or shock explanation of the 
S-shaped nuclear feature detected with {\it Chandra} suggest a fairly recent nuclear 
outburst, now spent. However, the present data are not
deep enough to either allow a detailed study of the morphology and spectral parameters 
of this feature, or a model-independent detection of the circum-nuclear hot ISM. A 
significantly deeper {\it Chandra} exposure will be needed to answer the many questions 
raised by the present data. Deeper VLA data are also needed to set a tighter 
constraint to the jet scenario.

\acknowledgments

We thank the {\it Chandra} X-ray Center DS and SDS teams for their efforts in reducing 
the data and developing the software used for the data reduction (SDP) and
analysis (CIAO). We have used the NASA funded services NED and ADS, and browsed 
the Hubble archive.
This work was supported by NASA contract NAS~8--39073 (CXC) and NASA grant GO3-4133X.



\begin{figure}
\caption{True color adaptively smoothed image of NGC~821. The ellipse marks the 25th 
mag 
isophote.}\label{}
\end{figure}

\begin{figure}
\caption{a) True color image of the central region of NGC~821, unsmoothed. 
The cross and surrounding circle represent the 2MASS nuclear position and 
uncertainty, from
NED. b) A larger 
field image  showing both the $wavdetect$ source regions (yellow) and the spectral 
extraction regions for the `jet' and the diffuse emission (light 
blue).}\label{fg:diff_img}
\end{figure}

\begin{figure}
\plotone{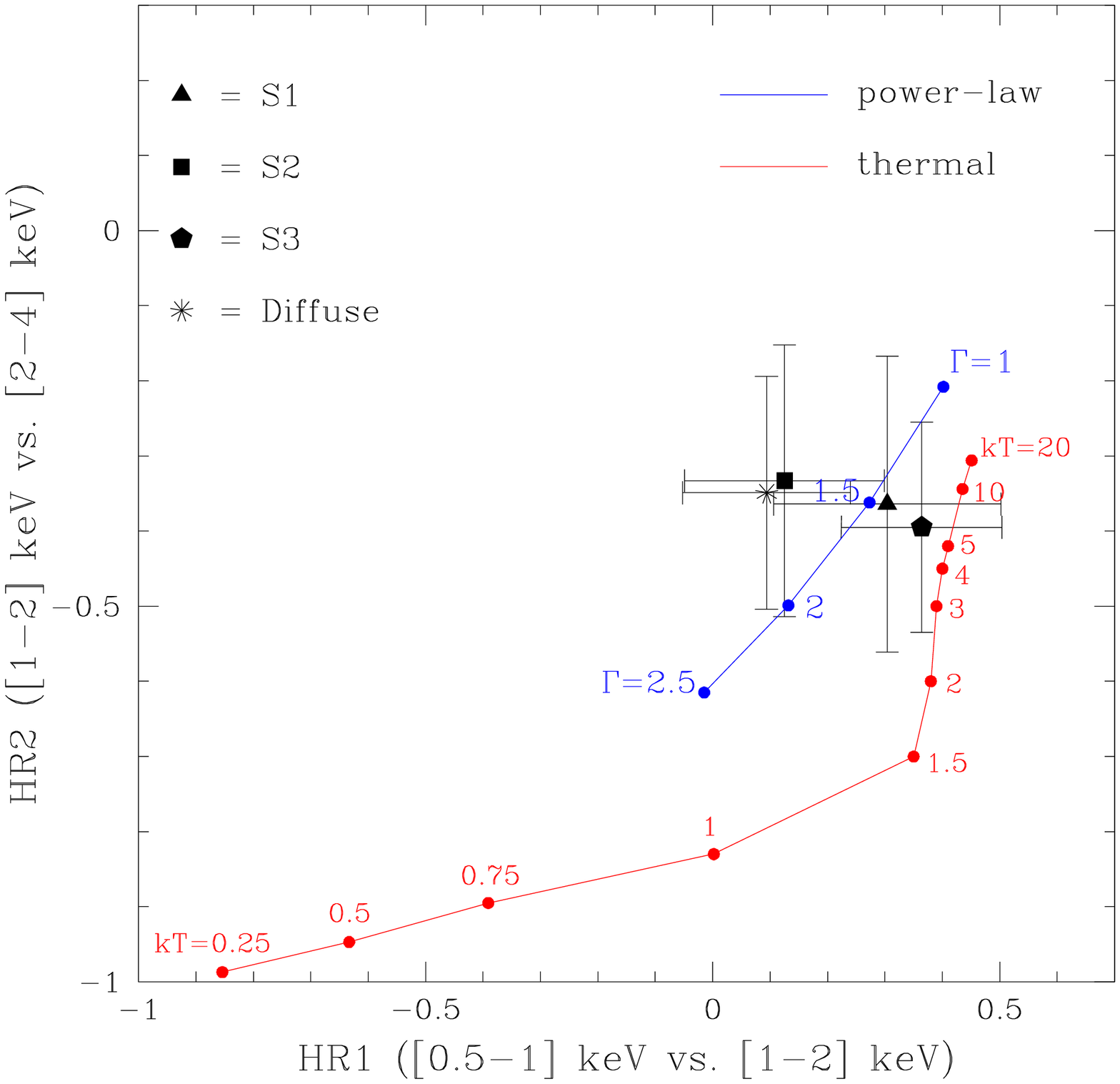}
\caption{X-ray colour-colour diagram of `jet' and diffuse emission regions. Typical
hardness ratios for a power-law model ($\Gamma=1-2.5$) are plotted in blue. Typical
hardness ratios for a thermal model ($kT=0.25-20$ keV) are plotted in red. Galactic
line-of-sight absorption ($N_H = 6.4 \times 10^{20} \rm cm^{-2}$, Stark et al 1992)
is assumed in both models.}\label{hratios}
\end{figure}

\clearpage

\begin{thebibliography}{}

\bibitem[Anders \& Grevesse 1989]{836} Anders, F. \& Grevesse, N. 1989,
Geochimica et Cosmochimica Acta, 53, 197

\bibitem[Arnaud 96]{839} Arnaud, K.A., 1996, Astronomical Data
Analysis Software and  Systems V, eds. Jacoby G. and Barnes J., ASP
Conf. Series vol. 101 

\bibitem[]{} Baganoff et al.\ 2001,
Nature, 413, 45

\bibitem[]{} Beuing, J., Döbereiner, S., Böhringer, H. \& Bender, R. 1999,
MNRAS, 302, 209

\bibitem[]{} Binney, J. \& Tabor, G. 1995, MNRAS, 276, 663



\bibitem[]{} Biretta, J.A., Zhou, F. \& Owen, F.N., 1995, ApJ, 447, 582 

\bibitem[]{} Biretta, J.A., Sparks, W.B. \& Macchetto, F., 1999, ApJ, 520, 621 

\bibitem[]{} Burbidge, G. R.  1959, ApJ, 129, 849

\bibitem[]{} Ciotti, L. \& Ostriker, J. P. 2001, ApJ, 551, 131

\bibitem[]{} Ciotti, L., Pellegrini, S., Renzini, A. \& D'Ercole, A. 1991,
ApJ, 376, 380

\bibitem[]{} De Young D., 1993, ApJ 402, 95.


 
\bibitem[]{} de Zeeuw, P. T. et al. 2002 MNRAS, 329, 513

\bibitem[]{} Di Matteo, T., Allen, S. W., Fabian, A. C., Wilson, A. S. \& Young, A. 
J. 2003 ApJ, 582, 133

\bibitem[]{} Di Matteo, T., Carilli, C. L., Fabian, A. C. 2001, ApJ, 547, 731

\bibitem[]{} Fabbiano, G. 1989, ARAA, 27, 87

\bibitem[]{} Fabbiano, G.  \& Juda, J. 1997, ApJ, 476, 666

\bibitem[]{} Fabbiano, G., Elvis, M., Markoff, S., Siemiginowska, A., Pellegrini, 
S., Zezas, A., Nicastro, F., Trinchieri, G.\& McDowell, J. 2003, ApJ, 588, 175

\bibitem[]{} Feigelson E.D., Schreier E.J., Delvaille J.P., Giacconi
  R., Grindlay J.E., \& Lightman A.P., 1981, ApJ, 251, 31.

\bibitem[]{} Felten, J. E. \&  Morrison, P. 1966 ApJ, 146, 686


\bibitem[]{} Ferrarese, L. \& Merritt, D. 2000, ApJ 539, L9

\bibitem[]{} Frank, J., King, A. \& Raine, D. J. 2002 Accretion Power in Astrophysics: 
Third Edition, [Cambridge: Cambridge University Press]

\bibitem[]{} Georgakakis, A., Hopkins, A. M., Caulton, A., Wiklind, T., Terlevich, 
A. I. \& Forbes, D. A. 2001, MNRAS, 326, 1431

\bibitem[]{} Gebhardt, K. et al. 2000, ApJ Letters, 539, L13

\bibitem[]{} Gebhardt, K. et al. 2003, ApJ,  583, 92

\bibitem[]{} Graham, A. W., Erwin, P., Caon, N., Trujillo, I. 2001, ApJ, 563, L11 

\bibitem[]{} Irwin J.A., \& Sarazin C.L. 1998, ApJ, 499, 650

\bibitem[]{} Janiuk, A., Czerny, B., Siemiginowska, A. \& Szczerba, R. 2004,
\apj, 602, 595

\bibitem[]{} Jones, C., Forman, W., Vikhlinin, A., Markevitch, M., David, L., 
Warmflash, A., Murray, S. \&  Nulsen, P. E. J. 2002, ApJ Letters, 567, L115

\bibitem[]{} Ho, L. C. 2002, ApJ, 564, 120

\bibitem[]{} Ho, L. C., Feigelson, E. D., Townsley, L. K., Sambruna, R. M., Garmire, 
Go. P., Brandt, W. N., Filippenko, A. V., Griffiths, R. E., Ptak, A. F., Sargent, W. 
L. W. 2001, ApJ 549, L51

\bibitem[]{} Ho, L. C., Filippenko, A. V., \& Sargent, W. L. W. 2003, ApJ, 583, 159

\bibitem[]{} Kadler,  M., Kerp, J.,  Ros, E., Falcke, H., Pogge, R. W., \& Zensus, 
J.A. 2004, A\&A, in press  (astro-ph/0403165)

\bibitem[]{} Kim, D.-W. \& Fabbiano, G. 2003  ApJ, 586, 826

\bibitem[]{} Kim, D.-W. \& Fabbiano, G. 2004, \apj, in press (astro-ph/0312104) 

\bibitem[]{} Kim, D.-W., Fabbiano, G. \& Mackie, G. 1998, ApJ, 497, 699

\bibitem[]{} Knapp, G. R., Faber, S. M. \& Gallagher, J. S. 1978, AJ, 83, 139

\bibitem[]{} Krolik J.H., 1999, {\em Active Galactic Nuclei}
  [Princeton:Princeton University Press].


\bibitem[]{} Leahy, J. P. 1991 in Beams and Jets in astrophysics, Edited by P.A. 
Hughes. Cambridge Astrophysics Series, No. 19, [Cambridge: Cambridge University 
Press], p. 100

\bibitem[]{} Loewenstein, M., Mushotzky, R. F., Angelini, L., Arnaud, K. A. \& 
Quataert, E.  2001, ApJ Letters, 555, L21

\bibitem[]{} Magorrian, J., Tremaine, S., Richstone, D., Bender, R., Bower, G., 
Dressler, A., Faber, S. M., Gebhardt, K., Green, R. \& Grillmair, C. 1998, AJ, 115, 
2285

\bibitem[]{} Markoff, S., Falcke, H., Yuan, F. \& Biermann, P. L. 2001, A\&A, 379, L13

\bibitem[]{} Martini, P., Pogge, R. W., Ravindranath, S. \& An, J. H. 2001, ApJ 562, 
139

\bibitem[]{} Narayan 2002, astro-ph/0201260

\bibitem[]{} Omma, H., Binney, J., Bryan, G. \& Slyz, A. 2004, MNRAS, 348, 1105

\bibitem[]{} Pellegrini S., \& Fabbiano G. 1994, ApJ, 429, 105

\bibitem[]{} Pellegrini, S., Baldi, A., Fabbiano, G. \&  Kim, D.-W. 2003a, ApJ, 597, 
175

\bibitem[]{} Pellegrini, S., Venturi, T., Comastri, A., Fabbiano, G., Fiore, F., 
Vignali, C., Morganti, R. \& Trinchieri, G. 2003b, ApJ, 585, 677

\bibitem[]{} Peterson B.M., 1997, {\em An Introduction to Active
  Galactic Nuclei} [Cambridge: Cambridge University Press].


\bibitem[]{} Phinney 1983, PhD thesis, University of Cambridge, UK

\bibitem[]{} Prugniel, P \& Simien, F. 1996 A\&A, 309, 749

\bibitem[]{} Ravindranath, S., Ho, L. C., Peng, C. Y., Filippenko, A. V. \& Sargent, 
W. L. W. 2001, AJ, 122, 653

\bibitem[]{} Rees, M. J. 1984 ARA\&A, 22, 471

\bibitem[]{} Reynolds, C.~S. 1997, PhD thesis, University of Cambridge, UK

\bibitem[]{} Reynolds, C.~S., Fabian, A.~C., Celotti, A., \& Rees, M.~J.\
1996, MNRAS, 283, 873

\bibitem[]{} Richstone, D. et al 1998, Nature, 395, 14 

\bibitem[]{} Rosati, P.  et al 2002, ApJ, 566, 667

\bibitem[]{} Schwartz D.A., Marshall H.L., Miller B.P., Worrall
D.M., Birkinshaw M., Lovell J.E.J., Jauncey D.L., Perlman E.S., Murphy
D.W., \& Preston, R. A., 2003, New AR, 47, 462

\bibitem[]{} Siemiginowska, A., Bechtold, J., Aldcroft, T. L., 
Elvis, M., Harris, D. E., Dobrzycki, A. 2002, \apj, 570. 543

\bibitem[]{} Siemiginowska, A., Czerny, B. \& Kostyunin, V.
1996, \apj, 458, 491 

\bibitem[]{} Siemiginowska,  A., Smith, R.~K., Aldcroft, T.~L.,
Schwartz, D.~A., Paerels, F., \& Petric, A.~O.\ 2003a, \apjl, 598, L15

\bibitem[]{} Siemiginowska,  A., Stanghellini, Carlo; Brunetti,
Gianfranco; Fiore F., Aldcroft T.L., Bechtold J., Elvis M., Murray
S.S., Antonelli L.A., \& Colafrancesco S., 2003b, \apj, 595, 643.

\bibitem[]{} Sambruna R.M., Gambill J.K., Maraschi L., Tavecchio F.,
  Cerutti R., Cheung C.C., Urry C.M. \& Chartas G., 2004, ApJ in
  press. astro-ph/0401475.
  
\bibitem[]{} Smith, R. K., Brickhouse, N. S., Liedahl, D. 
A., Raymond, J. C. 2001, ApJ, 556, L91

\bibitem[]{} Stark, A. A., Gammie, C. F., Wilson, R. W., Bally, J., Linke, R. A., 
Heiles, C. \& Hurwitz, M. 1992, ApJS, 79, 77

\bibitem[]{} van der Marel, R. P. 1999, AJ, 117, 744
 
\bibitem[Weisskopf \etal\ 2000]{939} Weisskopf, M., Tananbaum, H., Van
Speybroeck, L. \&  O'Dell, S., 2000, Proc. SPIE 4012, p. 2 (astro-ph 0004127)

\bibitem[]{} Wilson A.S., \& Yang Y., 2002, ApJ, 568, 133 
 
\end{thebibliography}
\end{document}